# Emirati-Accented Speaker Identification in Stressful Talking Conditions

Ismail Shahin[1], Ali Bou Nassif[2]
Department of Electrical and Computer Engineering
University of Sharjah
Sharjah, United Arab Emirates
[1]ismail@sharjah.ac.ae,
[2]anassif@sharjah.ac.ae

*Abstract*—This research is dedicated to improving "text-independent Emirati-accented speaker identification performance in stressful talking conditions" using three distinct classifiers: "First-Order Hidden Markov Models (HMM1s), Second-Order Hidden Markov Models (HMM2s), and Third-Order Hidden Markov Models (HMM3s)". The database that has been used in this work was collected from 25 per gender Emirati native speakers uttering eight widespread Emirati sentences in each of neutral, shouted, slow, loud, soft, and fast talking conditions. The extracted features of the captured database are called "Mel-Frequency Cepstral Coefficients (MFCCs)". Based on HMM1s, HMM2s, and HMM3s, average Emirati-accented "speaker identification accuracy in stressful conditions" is 58.6%, 61.1%, and 65.0%, respectively. The achieved "average speaker identification accuracy in stressful conditions based on HMM3s" is so similar to that attained in "subjective assessment by human listeners".

## I. Introduction and Literature Review

Speaker identification (SI) and speaker verification (SV) are two types of speaker recognition (SR). SI type is defined as the process of identifying the unknown speaker based on his/her voice, while SV is defined as the method of verifying (accepting or rejecting) the claimed speaker based on his /her voice. SI and SV have many applications in our daily life such as in the investigation of criminals to get the speculated suspects who uttered a voice produced at the occurrence of a crime and in confidential applications [1]. SR is clustered, based on the spoken text, into "text-dependent and text-independent kinds. In the text-dependent kind, SR requires the speaker to produce speech of the same text in both training and testing stages; on the other hand, the text-independent kind", SR is independent on the text being spoken.

Arabs in the Arab world can, generally, communicate among themselves in any of the four regional dialects of the Arabic language. "Egyptian (e.g. Egyptian), Levantine (e.g. Palestinian), North African (e.g. Tunisian), and Gulf Arabic (e.g. Emirati)" are the four regional dialects [2].

In this decade, there are growing number of studies in the areas of speech, speaker, and emotion recognition that utilize "Emirati-accented" speech database [3], [4], [5], [6], [7], [8].

Shahin and Ba-Hutair [3] studied "Emirati-accented speaker identification systems in a neutral talking condition based on Vector Quantization (VQ), Gaussian Mixture Models (GMMs), and Hidden Markov Models (HMMs) as classifiers. The Emirati dataset is made up of 25 men and 25 women Emirati native speakers. These speakers uttered 8 common Emirati sentences that are widely utilized in the United Arab Emirates society. Mel-Frequency Cepstral Coefficients (MFCCs) have been used as the extracted features of their dataset. Their results showed that VQ is superior to GMMs and HMMs for both text-dependent and text-independent cases".

Shahin [4] studied and evaluated a "text-independent speaker verification using Emirati corpus captured in a neutral talking environment. The corpus was portrayed from 25 male and 25 female Emirati local speakers who portrayed 8 frequently-used Emirati sentences. They extracted features using MFCCs of speech signals. Three different classifiers have been implemented in his research: First-Order Hidden Markov Models (HMM1s), Second-Order Hidden Markov Models (HMM2s), and Third-Order Hidden Markov Models (HMM3s). His study demonstrated that HMM3s lead HMM1s and HMM2s for a text-independent Emirati-accented speaker verification" in a neutral environment.

Further, Shahin *et.al* [5], captured "Emirati-accented speech dataset in each of neutral and shouted conditions to study and improve text-independent Emirati-accented speaker identification accuracy in shouted environment based on First-Order Circular Suprasegmental Hidden Markov Models (CSPHMM1s), Second-Order Circular Suprasegmental Hidden Markov Models (CSPHMM2s), and Third-Order Circular Suprasegmental Hidden Markov Models (CSPHMM3s) as classifiers. Their database was gathered from 50 Emirati national speakers (25 per gender) speaking 8 well-known Emirati sentences in each of neutral and shouted environments. The extracted features of their collected dataset are MFCCs. Their results gave average Emirati-accented speaker identification accuracy in neutral environment 94.0%, 95.2%, and 95.9% based on CSPHMM1s, CSPHMM2s, and CSPHMM3s, respectively. In contrast, the average accuracy in shouted environment is 51.3%, 55.5%, and 59.3% based, respectively, on CSPHMM1s, CSPHMM2s, and CSPHMM3s" [5].

Shahin targeted in one of his work [6] enhancing "text-independent Emirati-accented speaker identification accuracy in emotional talking environments based on three distinct classifiers: HMM1s, HMM2s, and HMM3s. The database was obtained from 50 Emirati native speakers (25 per gender) speaking 8 ordinary Emirati sentences in each of neutral, angry, sad, happy, disgust, and fear emotions. The extracted features of the database are MFCCs. His results yielded average Emirati-accented speaker identification accuracy in emotional environments 58.8%, 61.8%, and 65.9% based on HMM1s, HMM2s, and HMM3s", respectively [6].

In a recent study by Shahin *et.al* [7], they proposed an efficient methodology to improve "text-independent speaker identification accuracy in emotional talking environments based on a novel classifier called cascaded Gaussian Mixture Model-Deep Neural Network (GMM-DNN). Specifically, their work focused on proposing, implementing, and assessing a new framework for speaker identification in emotional talking environments based on cascaded Gaussian Mixture Model-Deep Neural Network as a classifier. Their results pointed out that the cascaded GMM-DNN classifier enhances speaker identification accuracy at diverse emotions using two different speech databases: Emirati speech database and Speech Under Simulated and Actual Stress (SUSAS) English dataset. The proposed classifier outperforms classical classifiers such as Multilayer Perceptron (MLP) and Support Vector Machine (SVM)" in each dataset [7].

In one of their most recent studies, Shahin *et.al* [8] aimed at recognizing emotions for a "text-independent and speaker-independent emotion recognition system based on an innovative classifier which is a mixture of sequential Gaussian Mixture Model and Deep Neural Network (GMM-DNN). This hybrid classifier has been evaluated for emotion recognition on Emirati speech database with six distinct emotions. The GMM-DNN has been compared with SVMs and MLP and its accuracy reached 83.97% while the other two operate at 80.33% and 69.78% using SVMs and MLP, respectively. Their results demonstrated that the hybrid classifier significantly yields greater emotion recognition accuracy than SVMs and MLP classifiers".

This work targets studying and enhancing "text-independent Emirati-accented speaker identification accuracy in stressful talking conditions based on three different classifiers: HMM1s, HMM2s, and HMM3s. These classifiers are novel for Emirati-accented speaker identification in such talking conditions. In this research, our speech corpus was gathered from 50 Emirati local speakers (25 per gender) talking 8 famous Emirati sentences in each of neutral, shouted, slow, loud, soft, and fast talking conditions. MFCCs have been used as the extracted features of the collected corpus".

The rest of this paper is organized as follows: Section II gives the fundamentals of "HMM1s, HMM2s, and HMM3s". Section III explains the details of the collected dataset used in this work and the extracted features. Section IV discusses "speaker identification algorithm based on HMM1s, HMM2s, and HMM3s" and the experiments. Section V gives the results achieved in the current work and their discussion. Finally, Section VI concludes the remarks of this study.

## II. FUNDAMENTALS OF HMM1S, HMM2S, AND HMM3S

### A. First-Order Hidden Markov Models

"Usually, HMMs can be defined as occurring in one of the $N$ diverse states: 1, 2, 3,…, $N$, at any discrete time moment $t$. The given states are denoted as,

$$s = \{s_1, s_2, s_3, ..., s_N\}$$

which are creators of a state sequence $q_t$, where at any time $t$: q = {$q_1, q_2, ..., q_T$}. At any discrete time $t$, the model is in a state $q_t$. At the discrete time $t$, the model causes an arbitrary move to a state $q_{t+1}$. The state transition probability matrix $A$ determines the probability of the following transition between states [9], [10], [11],

$$A = [a_{ij}] \qquad i, j = 1, 2, ..., N$$

where $a_{ij}$ designates the transition probability from a state $i$ to a state $j$.

In HMM1s, the state sequence is a first-order Markov chain where the stochastic process is shaped in a 2-D matrix of a priori transition probabilities ($a_{ij}$) between states $s_i$ and $s_j$ where $a_{ij}$ are given as:

$$a_{ij} = \text{Prob}\left(q_t = s_j | q_{t-1} = s_i\right) \qquad (1)$$

In such acoustic models, it is supposed that the state-transition probability at time $t+1$ relies only on the state of the Markov chain at time $t$. Readers can attain further information about HMM1s from references [9], [10].

### B. Second-Order Hidden Markov Models

The state sequence in HMM2s is a second-order Markov chain where the stochastic process is stated by a 3-D matrix ($a_{ijk}$). Hence, the transition probabilities in HMM2s are given as [12], [13]:

$$a_{ijk} = \text{Prob}\left(q_t = s_k | q_{t-1} = s_j, q_{t-2} = s_i\right) \qquad (2)$$

with the constraints,

$$\sum_{k=1}^{N} a_{ijk} = 1 \qquad N \geq i, j \geq 1$$

The state-transition probability in HMM2s at time $t+1$ depends on the states of the Markov chain at times $t$ and $t-1$. Additional information about HMM2s can be attained from references [12], [13]".

### C. Third-Order Hidden Markov Models

In HMM3s, the underlying state sequence is a "third-order Markov chain where the stochastic process is specified by a 4-D matrix ($a_{ijkw}$). Thus, the transition probabilities in HMM3s can be obtained as [14],

$$a_{ijkw} = \text{Prob}\left(q_t = s_w | q_{t-1} = s_k, q_{t-2} = s_j, q_{t-3} = s_i\right) \qquad (3)$$

with the constraints,

$$\sum_{w=1}^{N} a_{ijkw} = 1 \qquad N \geq i, j, k \geq 1$$

The probability of the state sequence, $Q \triangleq q_1, q_2, ..., q_T$, is described as:

$$\text{Prob}(Q) = \Psi_{q_1} a_{q_1 q_2 q_3} \prod_{t=4}^{T} a_{q_{t-3} q_{t-2} q_{t-1} q_t} \qquad (4)$$

where $\Psi_i$ is the probability of a state $s_i$ at time $t = 1$ and $a_{ijk}$ is the probability of the transition from a state $s_i$ to a state $s_k$ at time $t = 3$.

Given a sequence of observed vectors, $O \triangleq O_1, O_2, ..., O_T$, the joint state-output probability is stated as:

$$\text{Prob}(Q, O | \lambda) = \Psi_{q_1} b_{q_1}(O_1) a_{q_1 q_2 q_3} b_{q_3}(O_3) \cdot \prod_{t=4}^{T} a_{q_{t-3} q_{t-2} q_{t-1} q_t} b_{q_t}(O_t) \quad (5)$$

Supplementary information about HMM3s can be found in reference [14]".

## III. EMIRATI-ACCENTED SPEECH CORPUS AND EXTRACTION OF FEATURES

### A. Captured Emirati-Accented Speech Corpus

"Twenty-five native Emirati speakers per gender extending from 14 to 55 years old portrayed the Emirati-accented speech corpus (Arabic database). Every speaker spoke 8 familiar Emirati sentences that are regularly used in the United Arab Emirates society. The eight sentences were uttered by every speaker in each of neutral, shouted, slow, loud, soft, and fast talking conditions 9 times with a span of 2 – 5 seconds. The speakers were inexperienced to produce the Emirati sentences beforehand to prevent any overemphasized expressions (to make the dataset spontaneous). The overall number of recorded utterances was 12,600 ((50 speakers × first 4 sentences × 9 replicates/sentence in neutral environment for training session) + (50 speakers × last 4 sentences × 9 replications/sentence × 6 talking conditions for testing session)). The sentences are tabulated in Table I (the right column shows the sentences in Emirati accent, the left column displays the English version, and the middle column demonstrates the phonetic transcriptions of these sentences). This corpus was recorded in two isolated and different sessions: training session and testing session. The captured dataset was recorded in a clean environment in the College of Communication, University of Sharjah, United Arab Emirates by a set of specialized engineering students. The dataset was captured by a speech acquisition board using a 16-bit linear coding A/D converter and sampled at a sampling rate of 44.6 kHz. The signals were then down sampled to 12 kHz. The samples of signals were pre-emphasized and then segmented into frames of 20 ms each with 31.25% crossing between sequential frames".

### B. Extraction of Features

Mel-Frequency Cepstral Coefficients (MFCCs) "have been used in this study as the proper features that extract the phonetic content of Emirati-accented signals. Such features have been extensively used in many topics of speech. MFCCs have evidenced to outperform other coefficients and they have demonstrated to provide a high-level estimate of human auditory perception [15], [16], [17]. In this research, a 32-dimension feature analysis of MFCCs (16 static MFCCs and 16 delta MFCCs) was used to structure the observation vectors in each of HMM1s, HMM2s, and HMM3s. In every model, a continuous mixture observation density was selected with $N = 6$ states".

## IV. SPEAKER IDENTIFICATION ALGORITHM BASED ON HMM1S, HMM2S, AND HMM3S AND THE EXPERIMENTS

In each of "HMM1s, HMM2s, and HMM3s, the training phase (completely three independent training phases), the $v^{th}$ speaker model has been derived using the "first four sentences of the Emirati-accented speech corpus with 9 repetitions for each sentence portrayed by the $v^{th}$ speaker in neutral environment. The overall number of utterances that have been used to build the $v^{th}$ speaker model in each training phase is 36 (first 4 sentences × 9 times/sentence).

In the test (identification) phase of each of HMM1s, HMM2s, and HMM3s (totally three isolated test phases), each one of the fifty speakers individually portrays every sentence of the last four sentences of the corpus (text-independent) with 9 times/sentence in each of neutral, shouted, slow, loud, soft, and fast talking conditions. The entire number of utterances that have been used in each identification phase/talking condition is 1800 (50 speakers × last 4 sentences × 9 times/sentence). The probability of producing each utterance/speaker is independently calculated based on each of HMM1s, HMM2s, and HMM3s. For each one of these three classifiers, the model with the highest probability is chosen as the output of speaker identification as specified in the coming formula for each talking condition:

$$V^* = \arg\max_{50 \geq v \geq 1} \left\{ P\left(O | \lambda^v_{model}\right) \right\} \quad (6)$$

where $O$ is the observation vector that corresponds to the unknown speaker and $\lambda^v_{model}$ is the hidden Markov model (this model can be one of: HMM1s, HMM2s, or HMM3s) of the $v^{th}$ speaker".

## V. RESULTS AND DISCUSSION

Speaker identification accuracy in each of "neutral, shouted, slow, loud, soft, and fast" talking conditions using the "Emirati-accented speech dataset" based on "HMM1s, HMM2s, and HMM3s" as classifiers is given in Table II. It is very clear from this table that speaker identification accuracy is very high when speakers speak neutrally based on the three classifiers. On the other hand, the accuracy has been steeply declined when speakers speak in stressful conditions. Based on "HMM1s, HMM2s, and HMM3s", the table shows average speaker identification accuracy of 58.6%, 61.1%, and 65.0%, respectively. It is apparent from this table that "HMM3s" are superior to each of "HMM1s and HMM2s" in such conditions by 10.9% and 6.4%, respectively. It is apparent from this table that speaker identification accuracy is very high in neutral condition based on the three classifiers; however, the accuracy has been sharply declined in stressful conditions. This sharp decline comes from the mismatch that exists between the "training session in neutral environment" and the "testing session in stressful conditions". This mismatch negatively impacts "speaker identification accuracy" in stressful conditions.

A "statistical significance test" has been accomplished to show whether "speaker identification accuracy" alterations ("speaker identification accuracy" based on HMM3s and that based on each of HMM1s and HMM2s in stressful conditions) are real or only come from statistical variations.

The "statistical significance test" has been executed based on the "Student's *t* Distribution test" as presented by the following formula,

$$t_{\text{model 1, model 2}} = \frac{\overline{x}_{\text{model 1}} - \overline{x}_{\text{model 2}}}{SD_{\text{pooled}}} \quad (7)$$

where "$\overline{x}_{\text{model 1}}$ is the mean of the first sample (model 1) of size *n*, $\overline{x}_{\text{model 2}}$ is the mean of the second sample (model 2) of equal size, and $SD_{\text{pooled}}$ is the pooled standard deviation of the two samples (models)" given as,

$$SD_{\text{pooled}} = \sqrt{\frac{SD_{\text{model 1}}^2 + SD_{\text{model 2}}^2}{2}} \quad (8)$$

where "$SD_{\text{model 1}}$ is the standard deviation of the first sample (model 1) of size *n* and $SD_{\text{model 2}}$ is the standard deviation of the second sample (model 2) of equal size".

Table III demonstrates the "calculated *t* values" between "HMM3s" and each of "HMM1s and HMM2s" in "stressful conditions" using the "Emirati-accented" corpus. This table apprently tells that the "calculated *t* values" between "HMM3s" and every one of "HMM1s and HMM2s" are greater than the "tabulated critical value $t_{0.05}$ = 1.645 at 0.05 significant level". So, "HMM3s" significantly outperform each of "HMM1s and HMM2s" in such conditions. It is clear that "HMM3s" are better models than each of "HMM1s and HMM2s for speaker identification" in stressful talking conditions because the "characteristics of HMM3s" are made up of the characteristics of both "HMM1s and HMM2s".

In order to comprehensively assess the achieved speaker identification accuracy in stressful conditions using "Emirati-accented" corpus based on HMM3s, two additional experiments have been independently conducted in this work:

1) Experiment 1: Speaker identification accuracy utilizing the "Emirati-accented" database based on "HMM3s" has been competed with that based on four different "state-of-the-art models and classifiers". The four models and classifiers are: "GMMs [18], SVM [19], Genetic Algorithm (GA) [20], and VQ" [21]. Speaker identification accuracy in stressful conditions using the "Emirati-accented" dataset based on "GMMs, SVM, GA, VQ, and HMM3s" is displayed in Table IV. This table yields average speaker identification accuracy 59.1%, 60.8%, 59.1%, and 58.7% based on, respectively, "GMMs, SVM, GA, VQ". It is clear from this table and Table II that HMM3s outperform "GMMs, SVM, GA, and VQ for Emirati-accented speaker identification in stressful conditions" by 10.0%, 6.9%, 10.0%, and 10.8%, respectively.

2) Experiment 2: Using the Emirati-accented corpus, an "informal subjective assessment of HMM3s" has been accomplished with ten "nonprofessional listeners (human judges)". A sum of 2,400 utterances (50 speakers × 8 sentences × 6 stressful conditions) have been used in such an assessment. Throughout the evaluation, each listener was individually asked to identify the undetermined speaker in every stressful condition for each test utterance. The "average speaker identification accuracy in stressful conditions" based on the subjective assessment is shown in Table V. These averages are very alike to the attained averages in this work based on HMM3s as shown in Table II.

## VI. Concluding Remarks

"Text-independent Emirati-accented speaker identification in stressful talking conditions" has been considered based on three distinct classifiers: "HMM1s, HMM2s, and HMM3s". Some concluding remarks can be drawn in this study. First, third-order hidden Markov models are superior to each of first-order and second-order ones. Second, "HMM3s" lead each of "GMMs, SVM, GA, and VQ for speaker identification in stressful conditions". Finally, speaker identification accuracy is almost ideal in neutral condition; however, the accuracy has been steeply deteriorated in stressful conditions.

There are some limitations in this work. Firstly, the captured database is restricted to a sum of fifty speakers. Secondly, our corpus is acted. Finally, MFCCs have been used in this study as the appropriate features that extract the phonetic content of our corpus.

Our plan for future is to collect a comprehensive "Emirati-accented speech database" by comprising more speakers. Furthermore, we intend to involve speakers from the seven emirates of the "UAE (Abu Dhabi, Dubai, Sharjah, Ajman, Umm al-Qaiwain, Ras al-Khaimah, and Fujairah)". Finally, we plan to utilize deep neural networks [22] as classifiers to improve "Emirati-accented speaker identification accuracy in stressful conditions". In addition, our plan is to study and investigate Emirati-accented speaker identification in biased stressful talking environments [23], [24].

## Acknowledgment

"The authors of this work wish to acknowledge University of Sharjah for funding their research through the competitive research project entitled Capturing, Studying, and Analyzing Arabic Emirati-Accented Speech Database in Stressful and Emotional Talking Environments for Different Applications, No. 1602040349-P".

Table I. Emirati-accented speech database in its: English version, phonetic transcriptions, and Emirati accent

| No. | English Version | Phonetic Transcriptions | Emirati Accent |
|---|---|---|---|
| 1. | We will meet with you in an hour. | / bintlɑ:ga wɪjɑ:k ʕugub sɑ:ʕah / | بنتلاقى وياك عقب ساعة |
| 2. | Go to my father he wants you. | /si:r ʕɪnd abu:jeh yibɑ:k / | سير عند ابويه يباك |
| 3. | Bring my cell phone from the room. | /hɑ:t tilɪfu:ni: mɪnɪl ḥɪjrah / | هات تيلفوني من الحجرة |
| 4. | I am busy now I will talk to you later. | / maʃɣɔ:ɫ(a) ʌḥi:n barɑmsɪk ʕʌb sɑ:ʕəh / | مشغول/مشغولة الحين برمسك عقب |
| 5. | Every seller praises his market. | / kɪl byaiʕ yɪmdeḥ su:gah / | كل بياع يمدح سوقه |
| 6. | A stranger is a wolf whose bite wounds won't heal. | / ɪlġari:b ði:b w ʕaẓitah mɑṭi:b / | الغريب ذيب و عضته ما تطيب |
| 7. | Show respect around some people and show self-respect around other people. | / nɑ:æs ɪḥʃɪmhom w nɑ:s ɪḥʃɪm nafsak ʕanhom / | ناس احشمهم و ناس احشم نفسك عنهم |
| 8. | Don't criticize what you can't get and don't swirl around something you can't obtain. | / illi magdart tiyibah lɑ: tʕi:bah w illi mɑ:ṭu:lah lɑ: tḥu:m ḥu:lah / | اللي ما قدرت تييبه لا تعيبه و اللي ما تطوله لا تحوم حوله |

Table II. "Speaker identification accuracy in stressful talking conditions" using Emirati-accented corpus based on HMM1s, HMM2s, and HMM3s"

| Emotion | Speaker identification accuracy based on HMM1s (%) | | | Speaker identification accuracy based on HMM2s (%) | | | Speaker identification accuracy based on HMM3s (%) | | |
|---|---|---|---|---|---|---|---|---|---|
| | Males | Females | Average | Males | Females | Average | Males | Females | Average |
| Neutral | 95 | 93 | 94.0 | 96 | 94 | 95.0 | 96 | 96 | 96.0 |
| Shouted | 44 | 48 | 46.0 | 48 | 50 | 49.0 | 57 | 54 | 55.5 |
| Slow | 53 | 54 | 53.5 | 56 | 57 | 56.5 | 60 | 60 | 60.0 |
| Loud | 52 | 53 | 52.5 | 54 | 54 | 54.0 | 57 | 58 | 57.5 |
| Soft | 55 | 54 | 54.5 | 57 | 59 | 58.0 | 62 | 62 | 62.0 |
| Fast | 51 | 51 | 51.0 | 54 | 54 | 54.0 | 60 | 58 | 59.0 |

Table III. "Calculated *t* values between HMM3s and each of HMM1s and HMM2s in stressful talking conditions" using Emirati-accented corpus

| "t model 1, model 2" | Calculated *t* value |
|---|---|
| "t HMM3s, HMM1s" | 1.769 |
| "t HMM3s, HMM2s" | 1.701 |

Table IV
"Speaker identification accuracy in stressful talking conditions" using Emirati-accented corpus based on "GMMs, SVM, GA, VQ, and HMM3s"

| Emotion | Speaker identification accuracy (%) based on: | | | | |
|---|---|---|---|---|---|
| | GMMs | SVMs | GA | VQ | HMM3s |
| Neutral | 92.6 | 93.7 | 92.5 | 92.2 | 96.0 |
| Shouted | 50.4 | 53.4 | 50.7 | 51.0 | 55.5 |
| Slow | 54.3 | 56.4 | 54.4 | 53.0 | 60.0 |
| Loud | 50.7 | 52.0 | 50.5 | 50.4 | 57.5 |
| Soft | 52.8 | 54.1 | 53.0 | 52.2 | 62.0 |
| Fast | 53.9 | 55.2 | 53.6 | 53.1 | 59.0 |

Table V
Subjective assessment speaker identification accuracy

| Emotion | Speaker identification accuracy (%) |
|---|---|
| Neutral | 93.3 |
| Shouted | 52.9 |
| Slow | 63.8 |
| Loud | 59.8 |
| Soft | 64.7 |
| Fast | 62.5 |